\documentclass[letterpaper,british,prl,twocolumn]{revtex4}
\usepackage[T1]{fontenc}
\usepackage[latin9]{inputenc}
\usepackage{array}
\usepackage{booktabs}
\usepackage{multirow}
\usepackage{graphicx}

\makeatletter

\pdfpageheight\paperheight
\pdfpagewidth\paperwidth

\providecommand{\tabularnewline}{\\}

\@ifundefined{textcolor}{}
{%
 \definecolor{BLACK}{gray}{0}
 \definecolor{WHITE}{gray}{1}
 \definecolor{RED}{rgb}{1,0,0}
 \definecolor{GREEN}{rgb}{0,1,0}
 \definecolor{BLUE}{rgb}{0,0,1}
 \definecolor{CYAN}{cmyk}{1,0,0,0}
 \definecolor{MAGENTA}{cmyk}{0,1,0,0}
 \definecolor{YELLOW}{cmyk}{0,0,1,0}
}

\makeatother

\usepackage{babel}
\begin{document}

\title{Does coherence enhance transport in photosynthesis?}

\author{Ivan Kassal,\textsuperscript{1,2,3,}\footnote{Electronic address: \texttt{i.kassal@uq.edu.au}}
Joel Yuen-Zhou,\textsuperscript{3} and Saleh Rahimi-Keshari\textsuperscript{2,3}}

\affiliation{\textsuperscript{1}Centre for Engineered Quantum Systems, \textsuperscript{2}Centre
for Quantum Computing and Communication Technology, and \textsuperscript{3}School
of Mathematics and Physics, The University of Queensland, St Lucia
QLD 4072, Australia}

\date{\today}
\begin{abstract}
Recent observations of coherence in photosynthetic complexes have
led to the question of whether quantum effects can occur \emph{in
vivo}, not under femtosecond laser pulses but in incoherent sunlight
and at steady state, and, if so, whether the coherence explains the
high exciton transfer efficiency. We distinguish several types of
coherence and show that although some photosynthetic pathways are
partially coherent \emph{processes}, photosynthesis in nature proceeds
through stationary \emph{states}. This distinction allows us to rule
out several mechanisms of transport enhancement in sunlight. In particular,
although they are crucial for understanding exciton transport, neither
wavelike motion nor microscopic coherence, on their own, enhance the
efficiency. By contrast, two partially coherent mechanisms---ENAQT
and supertransfer---can enhance transport even in sunlight and thus
constitute motifs for the optimisation of artificial sunlight harvesting.
Finally, we clarify the importance of ultrafast spectroscopy in understanding
incoherent processes.
\end{abstract}
\maketitle
Recent observations of oscillatory spectroscopic signals in photosynthetic
light-harvesting complexes \cite{Engel:2007hb,Lee:2007hq,Collini:2010hb,Panitchayangkoon:2010fw,Wong:2012jd}
have led to suggestions that dynamical quantum effects may also occur
\emph{in vivo}, perhaps having a biological purpose \cite{Ishizaki:2009ky,Cheng:2009vb,Scholes:2011uj,Ishizaki:2011cx,Dawlaty:2012fs}
and having been favoured by natural selection. This question has been
difficult to answer because the strong, coherent laser light used
in experiments is substantially different from the weak, incoherent
sunlight. In particular, because sunlight intensity is constant on
excitonic timescales, photosynthetic light harvesting proceeds through
steady states and can be described by rate equations \cite{Mancal:2010kc}.
Indeed, the related problem of the photoisomerisation of rhodopsin---the
central event of vision---can be adequately described using a completely
incoherent model \cite{Hoki:2010wc}.

Determining whether coherence in photosynthetic complexes enhances
biological light harvesting requires distinguishing several types
of coherence, as we do in Sec. \ref{sec:Coherence-in-photosynthesis}
and Table \ref{tab:Types-of-coherence}. We find that although the
\emph{states} are static, the experiments have shown the \emph{process}
to be coherent.

Several mechanisms have been proposed by which coherence might enhance
the efficiency of exciton transport, and we discuss them in Sec. \ref{sec:Transport-efficiency},
ruling some out for photosynthesis in sunlight. For example, because
sunlight excites the entire complex---and not individual sites---and
because transport is through a steady state, there is no ``wavelike
transport'' that might speed up the exciton transfer. Similarly,
although individual microscopic realisations are more coherent than
the ensemble average, the microscopic coherences do not increase the
efficiency. Indeed, in Sec. \ref{sub:Spandrel} we argue that coherence
may be an evolutionary spandrel \cite{Gould:1979db} in most cases,
as it is quite likely that equally efficient incoherent transport
mechanisms are possible.

Nevertheless, we also identify mechanisms that \emph{can} enhance
transport. These mechanisms---ENAQT (Sec. \ref{sub:ENAQT}) and supertransfer
(Sec. \ref{sub:Supertransfer})---constitute viable design principles
for the engineering of artificial light-harvesting complexes.

Our findings do not imply that the oscillatory spectroscopic signals
with coherent light are irrelevant; quite the opposite, coherent optical
spectroscopy \cite{MukamelBook,minhaengbook} is indispensable for
elucidating transfer mechanisms and providing evidence of the strong
interchromophoric coupling that can lead to ENAQT and supertransfer
in nature.

\section{Coherence in photosynthesis\label{sec:Coherence-in-photosynthesis}}

Photosynthetic complexes consist of a number of (bacterio)chlorophyll
molecules, also called chromophores or sites, held in place by a protein
scaffold (see Fig. \ref{fig:Photosynthesis}) \cite{blankenship_molecular_2002}.
Each chlorophyll can be in the ground or excited states, and the question
of coherence in photosynthesis is, roughly speaking, whether a particular
excitation can be coherently delocalised over multiple sites. In the
following sections, we make this question more precise.

\begin{table*}
\begin{tabular}{>{\centering}p{1.6cm}>{\raggedright}p{2.5cm}>{\raggedright}p{5cm}>{\raggedright}p{7cm}}
\toprule 
\multicolumn{2}{c}{Type} & Definition & Remarks\tabularnewline
\midrule 
\multicolumn{2}{c}{Optical coherence} & Temporal correlation of the light field. & The coherence of the incident radiation affects the molecular states
that are created.\tabularnewline
\midrule
\multirow{4}{1.6cm}{State coherence} & Purity & $\mathrm{Tr}\left(\rho^{2}\right)$. & Basis-independent.\tabularnewline
\cmidrule{2-4} 
 & In a particular basis & Off-diagonal element of density matrix. & Basis-dependent. Coherence in energy basis required for isolated system
to undergo non-trivial time evolution. Coherence in site basis indicates
exciton delocalisation.\tabularnewline
\cmidrule{2-4} 
 & Static and dynamical coherence & At equilibrium or steady state, coherences are static (unchanging).
They are dynamical otherwise. & In unitary evolution, coherences in the energy basis evolve as $e^{-i\omega_{ij}t}$.
Wavelike transport requires dynamical coherence in the energy basis.
An open system may have static coherences in any basis.\tabularnewline
\cmidrule{2-4} 
 & Microscopic and ensemble coherence & Ensemble coherence is the expectation value of the energy-basis coherences
of each realisation (whose coherences are microscopic). & Because of averaging, ensemble coherence is less than the average
absolute value of microscopic coherences. Expectation values of observables
can be calculated either microscopically or using the ensemble average.\tabularnewline
\midrule
\multicolumn{2}{c}{Process coherence} & An open system evolves incoherently if the dissipation dominates unitary
evolution and partially coherently otherwise. & Basis-independent. Describes how long an initially coherent state
stays coherent. E.g., unitary evolution is coherent, Förster transfer
is incoherent. Importantly, a process can be coherent even if, in
particular cases, it proceeds through steady states (as in photosynthesis). \tabularnewline
\bottomrule
\end{tabular}

\caption{Types of coherence discussed in this work.\label{tab:Types-of-coherence}}
\end{table*}

\subsection{State and process coherence}

A quantum state, described by a density matrix $\rho$, is called
``pure'' if it can be represented by a wavefunction, $\rho=\left|\psi\right\rangle \left\langle \psi\right|$,
and ``mixed'' otherwise. The purity $\mathrm{Tr}\left(\rho^{2}\right)$
is a basis-independent measure of how close a state is to being pure.
Off-diagonal elements of $\rho$ are usually called ``coherences,''
but they are basis-dependent: a state diagonal in one orthonormal
basis will not be diagonal in any other. Two bases are particularly
important in discussing excitonic systems. The site basis is the basis
in which each exciton is localised on a particular site, while the
energy or exciton basis is the eigenbasis of the system Hamiltonian.
Because of the coupling between sites, the two bases usually do not
coincide.

Processes can also be described as coherent or incoherent. The evolution
of an open quantum system consists of a unitary part and a dissipative
part and the degree of process coherence depends on their relative
strengths \cite{May:2011vy}. For example, an isolated system evolving
unitarily is said to be evolving coherently, while Förster transfer
is incoherent because the donor and acceptor are each more strongly
coupled to their own dissipative environments then they are to each
other, meaning that the slow transport between them will proceed at
a simple rate \cite{May:2011vy}.

The distinction between state and process coherence is important.
Although long-lived state coherence implies partial process coherence,
the converse need not hold: a process can be coherent even if, in
particular cases, it proceeds through mixed states and can be described
by rate laws. The fact that the system \emph{would} evolve in a wavelike
fashion \emph{if} it were excited in the right way shows that the
process itself is coherent even if particular realisations are not.
That is, the coherent couplings are stronger than the noise even if
that can't be seen with, say, an initially mixed state.

\subsection{Excitation by lasers and by sunlight}

Until recently, most biological processes were thought to occur in
an environment so warm and wet that the dissipative term would dominate
the unitary term. The oscillations observed in coherent spectroscopic
experiments \cite{Engel:2007hb,Lee:2007hq,Collini:2010hb,Panitchayangkoon:2010fw,Wong:2012jd}
indicate that this need not be so. Femtosecond laser pulses can create
non-stationary states---states with coherences in the energy basis---and
the fact that these coherences persist for a long time is evidence
of partial process coherence. A substantial debate exists about the
extent to which these coherences are electronic or vibrational \cite{turner_vibrations,kaufmann,YuenZhou:2012hu},
but that distinction is beyond the scope of this work: we will consider
them as coherences between the vibronic eigenstates of the complete
molecular Hamiltonian.

While experiments have shown instances of photosynthetic light harvesting
to be partially coherent processes, to understand whether photosynthesis
involves coherent states in sunlight as well as under laser pulses,
we must consider optical coherence. Light can be coherent in different
ways, with various degrees of temporal, spectral, spatial, and polarisational
coherence \cite{Mandel:1995ub}. Spatial coherence can be assumed
because photosynthetic complexes are much smaller than visible wavelengths,
while polarisation coherence has little influence because of ensemble
orientational averaging. The two types of coherence that are more
important are the (related) temporal and spectral coherences. Classical
light is ``coherent'' if its phase can be predicted at all times,
and ``partially coherent'' otherwise. Sunlight is essentially incoherent
because for black-body radiation at 5870 K, the coherence time is
about 0.6 fs, shorter than other relevant timescales. Light can be
described in a fully quantised manner or semiclassically; the latter
generally suffices for biological problems.

The coherence of the light absorbed by a molecule affects the resulting
molecular state \cite{Scully:1997uy,Jiang:1991fk,Mancal:2010kc},
even if a single photon is absorbed \cite{Brumer:2011ty}. The effect
is most dramatic for isolated molecules: while coherent light can
excite coherences in the energy basis, an ensemble of isolated molecules
with non-degenerate levels, excited by incoherent light, will be in
a mixture of energy eigenstates \cite{Jiang:1991fk,Brumer:2011ty}.
Essentially, each non-degenerate transition will be driven by a different
frequency component in the light, and these will have uncorrelated
phases because the light is incoherent. When the random phases are
averaged, the coherences disappear whereas the populations do not.
Short pulses \emph{can} excite coherences \cite{Jiang:1991fk}, but
because sunlight intensity is constant on light-harvesting timescales,
these ``turn-on'' effects are of no concern. 

The behaviour is slightly different for open systems. Since the result
for isolated systems can be immediately applied to the system and
the bath as a whole, coherences between system-bath eigenstates must
vanish in the same way. However, upon tracing out the bath, energy-basis
coherences of the reduced system may be substantial and may even approach
the populations in magnitude \cite{Mancal:2010kc}. A recent calculation
of coherence dynamics in a multichromophoric system embedded in a
structured, non-Markovian environment, found the coherences to be
minimal \cite{Fassioli:2012gd}. 

Importantly, natural light harvesting occurs at steady state, because
the complexes are illuminated by weak sunlight whose intensity is
constant at all relevant timescales. The complex is essentially a
heat engine connected to two baths, the radiation at 5870 K and the
cooler surroundings, which extracts work while steadily transmitting
energy from the radiation to the surroundings. Therefore, whatever
coherences remain in the system after tracing out the bath will also
be \emph{stationary}, up to negligible turn-on transients; that is,
their phase will not evolve as $e^{-i\omega_{ij}t}$ as it would if
the system were isolated. In particular, this means that excitation
by sunlight will not be followed by excitonic wavelike motion. Large
stationary coherences can occur if the detuning between two levels
is small and their relaxation fast, meaning that they are close in
energy and broadened enough to make their spectral envelopes indistinguishable
\cite{Mancal:2010kc}. Indeed, tuneable systems could be arranged
so that the static state coherence between nearly degenerate levels
enhances the power from a photovoltaic cell \cite{Scully:2010cn,Scully:2011jj}. 

\begin{figure*}
\includegraphics[width=1\textwidth]{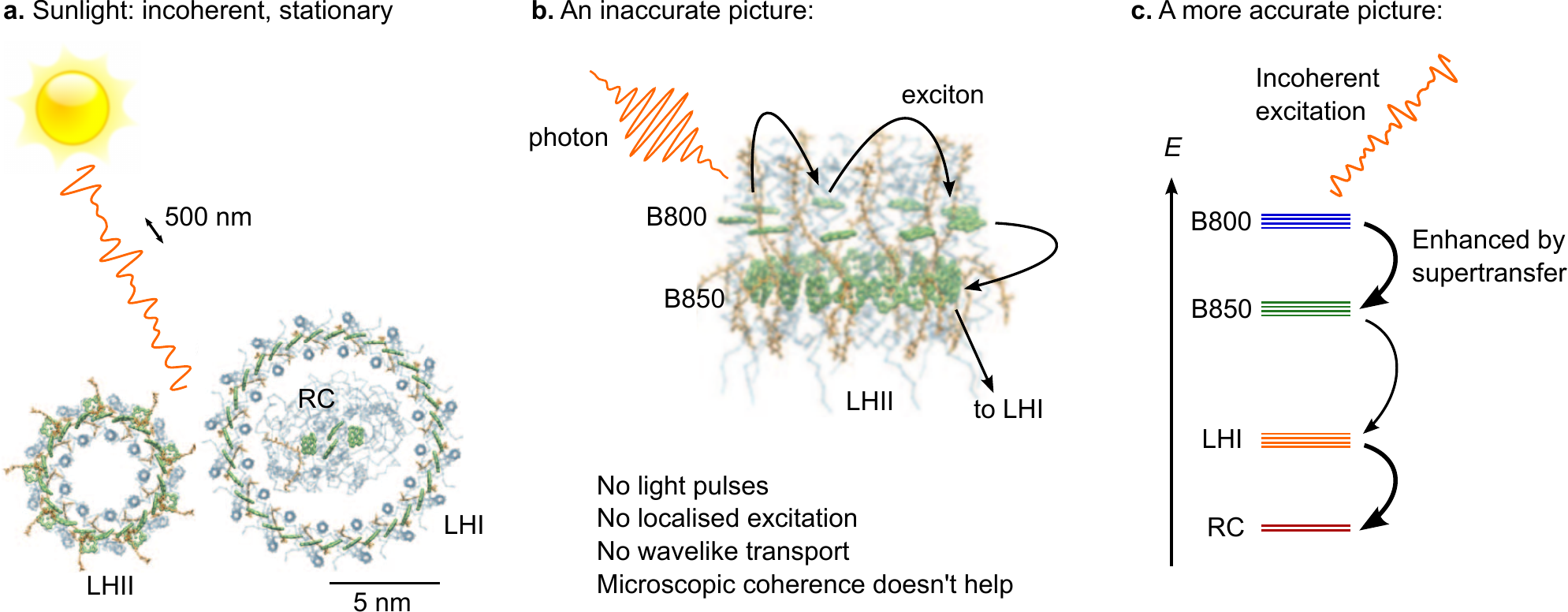}

\caption{Photosynthesis in incoherent light, illustrated with a simplified
model of LHII and LHI complexes from purple bacteria \cite{blankenship_molecular_2002}.
\textbf{a. }Sunlight, unlike femtosecond lasers, is incoherent and
stationary; therefore, photosynthesis operates at steady state. The
wavelength of light is much longer than the complex size. RC: reaction
centre. \textbf{b.} A common, but inaccurate view. Because the wavelengths
of light are long, it's not the case that a localised state is excited,
followed by wavelike transport. Consequently, several coherent mechanisms
are not relevant to photosynthesis \emph{in vivo}. \textbf{c.} Instead,
sunlight only excites stationary states that are diagonal or slightly
coherent in the energy basis. Incoherent long-range transfer between
different complexes (e.g., B800$\to$B850) can be enhanced by supertransfer,
a cooperative effect due to short-range process coherence (see Sec.
\ref{sub:Supertransfer}). Energies not to scale. Renderings of complexes
from \cite{Sener:2011ia}, with permission.\label{fig:Photosynthesis}}
\end{figure*}

\subsection{Absence of localised excitation}

Although much can be learned about transport by considering an exciton
initially localised on a particular site \cite{Mohseni:2008gp,Plenio:2008ff,Rebentrost:2009hu,Chin:2010hj,Hoyer:2010fl,Shabani:2012ir,Shim:2012el,Fujita:2012uo},
the picture changes when one models light absorption. The wavelength
of visible light is much larger than the size of photosynthetic complexes,
meaning that all the chromophores experience the same electric field.
Consequently, the steady state that is reached reflects the fact that
the incoming radiation excites collective eigenstates of the whole
complex, not individual sites.

The Fenna-Matthews-Olson complex (FMO) \cite{blankenship_molecular_2002}
has been modelled with the excitation starting on sites that are believed
to be closest to the chlorosome antenna that actually harvests the
light \cite{Mohseni:2008gp,Rebentrost:2009hu,Plenio:2008ff,Chin:2010hj,Hoyer:2010fl,Shabani:2012ir,Shim:2012el}.
This approach should be adapted for studying light harvesting in sunlight:
a chlorosome is as much an incoherent source of excitation as sunlight
is. The energy transfer time from chlorosome to FMO (more than 120
ps \cite{Psencik:2003bp}) is much larger than the chlorosome's dephasing
time, meaning that the transfer is by an incoherent Förster mechanism.
That is, the chlorosome excites FMO in much the same way as incoherent
light does (Sec. 4.3 of \cite{Mancal:2010kc}). Even if site 1 were
the only site coupled to the chlorosome, the chlorosome wouldn't coherently
excite site 1, to be followed by wavelike transport; rather, it would
excite a mixture of FMO eigenstates in proportion in which they are
found on site 1.

\subsection{Microscopic coherence\label{sub:Microscopic-coherence}}

Studies of photosynthetic complexes have usually considered ensembles.
In fact, each photosynthetic organism contains an ensemble of photosynthetic
complexes, each of which experiences a slightly different environment
(inhomogeneous disorder) \cite{Ishizaki:2011cx,Dawlaty:2012fs}. To
properly understand natural light harvesting, it will be important
to carry out single-molecule spectroscopic experiments to discern
the mechanistic details that may be washed out in the ensemble average
\cite{Dawlaty:2012fs,mukamel_rev_a,Lott04102011}.

An intriguing question is whether the state coherences of individual
ensemble realisations---the microscopic coherences---play a role.
Certainly, the ensemble density matrix, being the average of the microscopic
density matrices, will have smaller coherences than the average absolute
value of the microscopic coherences. Even if photosynthesis \emph{in
vivo} proceeded through diagonal steady states, the microscopic coherences
might be non-zero. It is tempting to think that the potentially large
microscopic coherences might affect light harvesting, by, for example,
increasing its efficiency. For example, if each realisation used state
coherence to enhance the efficiency, one would have to calculate the
efficiency for each realisation and \emph{then} average over the ensemble.

Having to simulate every realisation would be a Herculean task, but
is fortunately unnecessary. Changing the order of ensemble averaging
and calculating the expectation values of observables---including
the efficiency $\eta$---makes no difference:
\begin{equation}
\mathrm{Tr}\left(\eta\left\langle \rho_{S}\right\rangle _{\mathrm{ens}}\right)=\left\langle \mathrm{Tr}\left(\eta\,\rho_{S}\right)\right\rangle _{\mathrm{ens}},\label{eq:average}
\end{equation}
where $\left\langle \cdot\right\rangle _{\mathrm{ens}}$ is the average
over realisations $\rho_{S}$. Although some elements of the ensemble
might benefit from state coherence, the benefit will be compensated
in others. For example, if we imagine that the efficiency is proportional
to a coherence, $\eta\propto\rho_{12}$, then $\left\langle \eta\right\rangle _{\mathrm{ens}}\propto\left\langle \rho_{12}\right\rangle _{\mathrm{ens}}=0$
if the ensemble is diagonal. Although it is tempting to speculate
about the efficiency including terms like $|\rho_{12}|^{2}$, whose
average would not vanish, it should be remembered that $|\rho_{12}|^{2}$
is not a linear operator on $\rho$ and therefore not a valid observable.
Likewise, the purity $\mathrm{Tr}\left(\rho^{2}\right)$ is not an
observable, so there is no contradiction if an ensemble of pure states
has a purity less than 1. Non-linear expressions such as these are
not observables because they cannot be measured in a single-shot measurement.
Determining their expectation values is possible, but it requires
the experimental ability to produce multiple copies of the same quantum
state, which is not the situation in a genuinely random ensemble.
Additionally, we do not claim that quantum systems can only respond
linearly to external perturbations, but only that expectation values
of observables are linear in $\rho$ even if the response is nonlinear
in the perturbation. 

A similar situation occurs if one considers sunlight as a train of
femtosecond pulses \cite{Cheng:2009vb}, in which case there is an
ensemble of the phases and arrival times of these pulses. Even if
each pulse could excite non-stationary states like a femtosecond laser
could, the simpler, incoherent ensemble average will reproduce all
the observables. 

We do not seek to undermine the usefulness of single-molecule experiments,
which are indispensable in elucidating biological \emph{mechanisms}---as
patch-clamping was for ion channels. We merely stress that once the
mechanism is known, the\emph{ outcome} of a process can be calculated
either microscopically or using the ensemble average, indicating that
microscopic coherences do not play an important role in sunlight harvesting.

\section{Exciton transport efficiency\label{sec:Transport-efficiency}}

Because the process of light-harvesting is partially coherent, it
is tantalising to wonder whether the coherence might enhance the efficiency
of exciton transport, suggesting it was selected by natural selection.
The efficiency can be defined as the proportion of the initially created
excitons that reach the reaction centre. Several mechanisms have been
proposed by which coherence might enhance transport efficiency, and
we consider how each of them might operate in the steady-state regime
of natural light harvesting. We have already seen, in Sec. \ref{sub:Microscopic-coherence},
that microscopic coherence cannot enhance efficiency, and in Sec.
\ref{sub:Delocalisation} we rule out another mechanism. Nevertheless,
two mechanisms may be said to operate: ENAQT and supertransfer, discussed
in Sec. \ref{sub:ENAQT} and \ref{sub:Supertransfer}, respectively.
This is not to say that coherence is \emph{necessary} for highly efficient
transport---in Sec. \ref{sub:Spandrel} we argue that natural coherence
may be an evolutionary spandrel and that although ENAQT and supertransfer
may enhance the efficiency of artificial light-harvesting complexes,
the possibility of efficient incoherent transport should not be overlooked.

\subsection{Faster delocalisation\label{sub:Delocalisation}}

The simplest example of a difference between quantum and classical
transport occurs on ordered, infinite lattices. On a one-dimensional
lattice, quantum transport is ``ballistic'' because the variance
of the particle's wavefunction is proportional to time, $\Delta x_{\mathrm{quant}}=c_{\mathrm{quant}}t$.
Classical transport, say by random walk, is ``diffusive'', $\Delta x_{\mathrm{class}}=c_{\mathrm{class}}\sqrt{t}$.
Therefore, at sufficiently long times, $\Delta x_{\mathrm{quant}}$
will exceed $\Delta x_{\mathrm{class}}$, even if $c_{\mathrm{class}}>c_{\mathrm{quant}}$. 

If an excitation were initially localised, coherent delocalisation
might well enhance transport to a distant reaction centre. But as
we noted above, photosynthetic complexes are much smaller than the
wavelength of light, meaning that initial excitations are not localised.
Furthermore, in any \emph{finite} system, a classical particle could
spread over the entire complex faster than a quantum particle if $c_{\mathrm{class}}$
were sufficiently large. Therefore, even if an exciton in a biological
complex initially spread ballistically \cite{Hoyer:2010fl}, incoherent
transport could cause faster delocalisation if the incoherent transfer
rates were higher. In other words, the coherent speed-up of delocalisation
cannot be said to be responsible for the high transport efficiency.

\subsection{ENAQT\label{sub:ENAQT}}

Environment-assisted quantum transport (ENAQT) can occur in systems
whose evolution can be modified from coherent to incoherent using
an adjustable coupling to a particular bath. ENAQT occurs if the efficiency
of transport from one site to another is highest in the intermediate
coupling regime, i.e., higher than it would be in either the unitary
or incoherent limits \cite{Rebentrost:2009hu,Plenio:2008ff,Kassal:2012jd,Shabani:2012ir}.
Although previous work has considered initially localised excitations,
the steady state version is easily constructed as well. 

In disordered systems, initially localised excitations may be prevented
from delocalising by coherent effects such as Anderson localisation.
ENAQT occurs if moderate decoherence destroys the coherent localisation,
allowing the particle to reach its target. Similarly, very strong
decoherence can prevent transport, meaning that adding partial coherence
can optimise the efficiency. The extent of ENAQT is very dependent
on the nature of the particular bath being studied. For certain realistic
baths, the transport efficiency in FMO has been found to be optimised
in an intermediate coupling regime \cite{Shabani:2012ir}; therefore,
that complex can be said to have enhanced transport over the hypothetical
case of weaker or stronger bath coupling. However, as we argue in
Sec. \ref{sub:Spandrel}, this does not show that coherence is necessary
for the high efficiency.

\subsection{Supertransfer\label{sub:Supertransfer}}

The second design motif is supertransfer, an enhancement of long-range
incoherent transport by short-range process coherence \cite{Meier:1997fq,Lloyd:2010fz,Strumpfer:2012ep}.
Named after superradiance, it involves a donor complex and an acceptor
complex, each composed of several chromophores. The two complexes
are far apart and the weak transfer between them incoherent (Förster),
but the total incoherent rate depends on the process coherence \emph{within}
the donor. In the complete absence of process coherence, each chromophore
in the donor is independently incoherently coupled to each chromophore
in the acceptor by the dipole-dipole interaction. In the alternative
case, the excitons within the donor are delocalised across multiple
sites, allowing for cooperative transfer and an enhanced overall incoherent
rate. In excitonic systems this effect is also known as multi-chromophoric
Förster resonant energy transfer \cite{Jang:2004vk,Jang:2007cz}.

In the extreme case, incoherent transfer of excitons symmetrically
delocalised across $M$ chromophores on the donor can be up to $M$
times faster than if the chromophores only communicated individually
\cite{Lloyd:2010fz} (see also the renormalisation scheme in \cite{Ringsmuth:2012hb}).
For example, we consider a donor and an acceptor, each composed of
two chromophores, all of whose dipole moments are parallel and with
magnitude $\mu$. In the incoherent case, each chromophore in the
donor has a $\frac{1}{2}$ chance of being occupied and transmits
to each acceptor chromophore with a Fermi-golden-rule rate $\gamma\sim|\mu_{D}\mu_{A}|^{2}=\mu^{4}$.
This gives a total incoherent rate of $\Gamma_{\mathrm{incoh}}=2\gamma$.
If local coherence is present and the donor is in the symmetric ground
state with dipole $\left(\mu_{1}+\mu_{2}\right)/\sqrt{2}$ that communicates
with the corresponding state on the acceptor, the total incoherent
rate is doubled: 
\begin{equation}
\Gamma_{\mathrm{coh}}\sim\left|\left(\frac{\mu_{1}+\mu_{2}}{\sqrt{2}}\right)_{D}\left(\frac{\mu_{1}+\mu_{2}}{\sqrt{2}}\right)_{A}\right|^{2}=4\mu^{4}=2\Gamma_{\mathrm{incoh}}.
\end{equation}

Supertransfer persists in incoherent light and at steady state, meaning
that biomolecular networks in which supertransfer occurs, such as
the LHI and LHII complexes \cite{Jang:2007cz} in Fig. \ref{fig:Photosynthesis},
may be said to have enhanced efficiency compared to the situation
where all the chromophores were individually coupled \cite{Strumpfer:2012ep}.

\subsection{Is coherence a spandrel?\label{sub:Spandrel}}

ENAQT and supertransfer are both robust effects that may be used to
explain the high transport efficiency in some complexes and that may
be used in the design of artificial complexes. However, we must caution
against the conclusions that process coherence always assists transport
or that it is necessary for high transport efficiencies. Indeed, for
any particularly efficient coherent process, one could increase the
efficiency by adding additional incoherent rates direct to the reaction
centre. Therefore, in engineering artificial light-harvesting systems,
one should not consider process coherence necessarily advantageous.
This is not to say that there are no design motifs to be learned from
photosynthesis: in certain artificial systems, ENAQT and supertransfer
may be of great use. 

The same caution should be applied to hypotheses that coherence is
responsible for the high efficiency of photosynthetic exciton transport
or that it was favoured by natural selection. Even if, for a particular
bath, ENAQT calculations show that moderate coherence helps, we should
recognise that evolution was not constrained to any particular bath
(i.e., protein cage) and that a different bath could have yielded
a more efficient, albeit incoherent, process. For this reason, we
suspect that the observed process coherence may be an evolutionary
spandrel \cite{Gould:1979db}. It could well be that in trying to
increase incoherent couplings and create an energy funnel to the reaction
centre, evolution brought the chromophores closer, making strong interchromophoric
couplings---and thus partial process coherence---unavoidable.

\section{Conclusions}

Although natural light harvesting proceeds through stationary states,
long-lived dynamical coherences in ultrafast experiments remain remarkable
because they show that the couplings between the chromophores are
stronger than their couplings to their respective baths, which can
indicate ENAQT or supertransfer even under incoherent illumination.
Therefore, this artificial phenomenon is relevant not because it occurs
in the same way \emph{in vivo}, but because it may indicate energy
transport mechanisms that are qualitatively different from the previously
assumed incoherent site-to-site hopping. Because this is the case
if the observed dynamical coherences are of electronic and not vibrational
nature, additional studies should address the origin of the oscillations
\cite{kaufmann,YuenZhou:2012hu}.

A remaining challenge is to construct models of energy transfer under
incoherent light based on spectroscopic data acquired with coherent
light sources. Quantum process tomography helps achieve this goal
by systematically correlating the prepared input and measured output
states in the excitonic system \cite{YuenZhou:2011jt,YuenZhou:2011vt}.
A complete characterisation may permit the control of energy flow
using coherent light in the spirit of quantum control \cite{Herek:2002wy,Shapiro:2012ui},
in which case dynamical coherences would be essential to achieving
the desired goal.
\begin{acknowledgments}
We thank Alán Aspuru-Guzik, Alessandro Fedrizzi, Daniel James, Jacob
Krich, Gerard Milburn, Ben Powell, Tim Ralph, Drew Ringsmuth, Tom
Stace, and Andrew White for valuable discussions. This work was supported
by a UQ Postdoctoral Research Fellowship, a UQ Early Career Researcher
grant, and the Australian Research Council Centre of Excellence for
Quantum Computation and Communication Technology (project CE110001027).
\end{acknowledgments}

\end{document}